\documentclass[12pt]{article}
\usepackage{amssymb}
\usepackage{graphicx}
\usepackage{psfrag}
\usepackage{amsfonts}
\usepackage{epsfig}
\usepackage{rotate}
\usepackage{latexsym}
\newcommand{\sef}{\sin^2 \theta_{eff}^{lept}}
\newcommand{\ini}{\begin{equation}}
\newcommand{\fin}{\end{equation}}
\newcommand{\sms}{\hat{s}^2}
\newcommand{\cms}{\hat{c}^2}
\newcommand{\ems}{\hat{e}^2}
\newcommand{\es}{s_{eff}^2}

\newcommand{\msbar}{\overline{MS}}
\newcommand{\dah}{\Delta \alpha_h^{(5)}}
\newcommand{\dahmz}{\Delta \alpha_h^{\left( 5 \right)}(M_Z)}
\newcommand{\asmz}{\alpha_s \left( M_Z \right)}

\makeatletter
 \def\citenum#1{{\def\@cite##1##2{##1}\cite{#1}}}
\def\citea#1{\@cite{#1}{}}

\newcount\@tempcntc
\def\@citex[#1]#2{\if@filesw\immediate\write\@auxout{\string\citation{#2}}\fi
  \@tempcnta\z@\@tempcntb\m@ne\def\@citea{}\@cite{\@for\@citeb:=#2\do
    {\@ifundefined
       {b@\@citeb}{\@citeo\@tempcntb\m@ne\@citea\def\@citea{,}{\bf }\@warning
       {Citation `\@citeb' on page \thepage \space undefined}}%
    {\setbox\z@\hbox{\global\@tempcntc0\csname b@\@citeb\endcsname\relax}%
     \ifnum\@tempcntc=\z@ \@citeo\@tempcntb\m@ne
       \@citea\def\@citea{,}\hbox{\csname b@\@citeb\endcsname}%
     \else
      \advance\@tempcntb\@ne
      \ifnum\@tempcntb=\@tempcntc
      \else\advance\@tempcntb\m@ne\@citeo
      \@tempcnta\@tempcntc\@tempcntb\@tempcntc\fi\fi}}\@citeo}{#1}}
\def\@citeo{\ifnum\@tempcnta>\@tempcntb\else\@citea\def\@citea{,}%
  \ifnum\@tempcnta=\@tempcntb\the\@tempcnta\else
  {\advance\@tempcnta\@ne\ifnum\@tempcnta=\@tempcntb \else\def\@citea{--}\fi
    \advance\@tempcnta\m@ne\the\@tempcnta\@citea\the\@tempcntb}\fi\fi}
\makeatother

\begin{document}

\hyphenation{re-nor-ma-li-za-tion}

\begin{flushright}
NYU-TH/02-03-01\\
hep-ph/0203224
\end{flushright}

\vspace{0.5cm}
\begin{center}
\boldmath{\Large \bf Simple Formulae for $\sin^2 \!\theta_{eff}^{lept}$, $M_W$,
$\Gamma_l$, and their Physical Applications}\unboldmath\\
\vspace{0.2cm}
\vspace{0.5cm}
{\large 
A.~Ferroglia$^{a,}$\footnote{e-mail: andrea.ferroglia@physics.nyu.edu}, 
G.~Ossola$^{a,}$\footnote{e-mail: giovanni.ossola@physics.nyu.edu}, 
M.~Passera$^{b,}$\footnote{e-mail: passera@pd.infn.it}, 
and A.~Sirlin$^{a,}$\footnote{e-mail: alberto.sirlin@nyu.edu}}


\vspace{0.5cm}
$^a${\it Department of Physics, New York University,\\
4 Washington Place, New York, NY 10003, USA} \\
\vspace{0.5cm}
$^b${\it Dipartimento di Fisica ``G.~Galilei'', Universit\`{a} 
        di Padova and \\ INFN, Sezione di Padova, Via Marzolo 8,
        I-35131, Padova, Italy}
\end{center}
\bigskip

\begin{center}
\bf Abstract 
\end{center}
{ \small The calculation of the partial widths of the $Z$ boson in the 
effective scheme is discussed.  
Simple formulae for $\sef$, $M_W$, and  $\Gamma_l$ in the 
on-shell, $\msbar$, and effective renormalization schemes are
presented. Their
domain of validity extends to low $M_H$ values probed by 
current experiments, and to the range of $\dah$ results 
from recent calculations.
Physical applications of these formulae are illustrated.
A comparative analysis of the above mentioned schemes is presented.
The paper includes a discussion of the discrepancy of the $\sef$ values derived
from the leptonic and hadronic sectors and of the
shortcomings of theoretical error estimates of $\sef$ based 
solely on those affecting $M_W$.}

\newpage
 
\section{Introduction}

The on-shell \cite{c1,c2} and $\msbar$ 
formulations \cite{c3,c4,c5,c6,c7} are two 
of the most frequently employed schemes of renormalization in the evaluation  
of electroweak corrections. The on-shell approach (OS) is very ``physical'',
as it identifies the renormalized parameters with observable, scale-independent
constants, such as $\alpha$, $M_W$ and $M_Z$. On the other hand,
$\msbar$ calculations exhibit very good convergence properties. In fact,
they follow closely the structure of the unrenormalized theory and,
for this reason, avoid large finite corrections that are frequently induced 
by renormalization. They also involve inherently scale-dependent parameters,
such as $\hat{\alpha} \left( \mu  \right) $ 
and $\sin^2 \hat{\theta}_W \left(\mu \right)$, which play
a crucial role in the analysis of grand-unification.

Recently, a novel approach, the effective scheme of renormalization (EFF), 
has been proposed \cite{c8,c9}. It shares the good convergence properties
of the $\msbar$ formulation but, at the same time, eliminates the residual 
scale dependence induced, in finite orders, by the truncation of the 
perturbative series. Like the OS approach, it identifies the renormalized
parameters with observable, scale-independent constants, such as $\alpha$,
$\es \equiv \sin^2 \theta_{eff}^{lept}$, and $M_Z$.
Thus, in the OS, $\msbar$, and EFF schemes, the basic electroweak mixing
parameter is identified with  $\sin^2 \theta_W \equiv 1-M_W^2/M_Z^2$,  
 $\sin^2 \hat{\theta}_W \left(\mu \right)$, and $\es$, respectively.\\
It has been shown that the incorporation of the 
${\cal O}(\alpha^2 M_t^2/M_W^2)$ contributions \cite{c10} greatly reduces 
the scheme and scale dependence of the
radiative corrections in the theoretical evaluation of $\es$
and $M_W$, as well as the upper bound on $M_H$ \cite{c11,c12}. In these papers
the calculations were carried out in two different implementations of the
on-shell scheme of renormalization, denoted as OSI and OSII, as well as
in the $\msbar$ framework. Comparison of the three calculations led to
an analysis of the scheme dependence and, by inference, to an estimate of
the theoretical error due to the truncation of the perturbative series.
Ref.~\cite{c13} extended these studies to the analysis of the partial 
leptonic widths
$\Gamma_l$ of the $Z$ boson. In turn, Refs.~\cite{c8,c9} discussed the
results of the $\es$ and $M_W$ calculations in the EFF scheme.

One of the objectives of this paper is to present simple analytic formulae for
the precisely measured parameters $\es$, $M_W$, and $\Gamma_l$ in the four
schemes ($\msbar$, OSI, OSII, EFF), as functions of $M_H$ and important
inputs such as $M_t$, $\dahmz$, and $\alpha_s (M_Z)$, and to illustrate their
usefulness in physical applications. 
Expressions of this type
have been discussed before in the framework of the OS and $\msbar$
schemes \cite{c12,c13}. In the present paper we extend the range 
of validity of the formulae to lower values of $M_H$ probed by recent 
measurements, and include the corresponding expressions in the EFF scheme.

The plan of the paper is the following: Section 2 outlines the theoretical 
derivation of $\Gamma_l$ in the EFF scheme, which has not been discussed in
the literature. Section 3 presents simple formulae for
$\es$, $M_W$, and $\Gamma_l$  in the four schemes and shows how to apply 
them in order to obtain
information about $M_H$ and to predict $M_W$. A discussion of the
discrepancy of the $\es$ values derived from the leptonic and hadronic
sectors is included.  
Section 4 presents a comparative analysis of the results in the four schemes
and a discussion of the theoretical errors associated with the
$\es$ and $M_W$ calculations.

\boldmath
\section{Leptonic Partial Widths of the $Z$ boson in the EFF scheme}
\unboldmath
Theoretical calculations of the partial widths of the $Z$ boson into leptons
and the $u$, $d$, $s$, $c$ quark channels that incorporate the corrections of
order 
${\cal O}\left( g^4 (M_t^2/M_W^2)^n \right)$ $(n=1,2)$ have been carried out
in Ref.~\cite{c13} in the framework of the $\msbar$, OSI, and OSII schemes.
In particular, simple analytic formulae for the leptonic partial widths
$\Gamma_l$ have been presented in the same work.

In this Section we extend the analysis of $\Gamma_l$ to the EFF scheme. As
in Refs.~\cite{c8,c11,c13} we retain complete one-loop corrections and
two-loop effects enhanced by factors $\left( M_t^2/M_W^2 \right)^n$ $(n=1,2)$
or involving the product of one-loop imaginary parts. In the latter case, the 
rationale is that such products are quite sizable, relative to typical
${\cal O}(g^4)$ contributions, since imaginary parts involve several additive
terms.

It is convenient to start with the $\msbar$ expression \cite{c6,c13}:
\ini \label{e11}
\Gamma_Z^l = \frac{\ems}{\sms \cms} \frac{M_Z}{96 \pi} | \hat{\eta}_l |
\left\{ 1 - 4 |Q_l| \sms Re\ \hat{k}_l +8 Q^2_l \hat{s}^4 | \hat{k}_l |^2  \right\} 
+ {\cal O}\left( \frac{m_l^2}{M_Z^2} \right) \, ,
\fin
where $\ems$ and $\sms = 1 - \cms$ are abbreviations for the $\msbar$ parameters
$\ems(\mu)$ and  $\sin^2 \hat{\theta}_W \left(\mu \right)$, respectively,
$Q_l$ is the charge of the final lepton, $m_l$ is its mass, and $\hat{k}_l (\mu)$
and $\hat{\eta}_l (\mu)$ stand for appropriate electroweak form factors.
QED corrections are included by appending an overall factor of
$1 + 3 \alpha /4 \pi$.
An expression analogous to Eq.~(\ref{e11}) is valid for the four lightest
quarks provided a color factor $N_c =3$ is appended and
$Q_l \hat{k}_l$ is replaced by $Q_f \hat{k}_f$ ($f$=$u$-$c$). 
In this case, one-loop
QCD corrections are included by inserting a factor $1 + \alpha_s (M_z)/
4 \pi$.
These expressions are conventionally evaluated at $\mu = M_Z$ but,
in principle, they may be studied over a range of $\mu$-values since the
truncation of the perturbative series induces, in general, a residual scale
dependence.

As in Ref.~\cite{c8}, we attempt to express Eq.~(\ref{e11}) in terms of scale-independent
constants such as $G_{\mu}$, $M_Z$, and $\es$. Recalling the basic relation \cite{c7}
\ini \label{e12}
 \es = \sms  Re\ \hat{k}_l   \, ,
\fin
we note that, through terms of ${\cal O}(g^4)$, the expression between curly
brackets in Eq.~(\ref{e11}) becomes
\ini \label{e13}
  1 - 4 |Q_l| \sms Re\ \hat{k}_l +8 Q^2_l \hat{s}^4 | \hat{k}_l |^2 =
 1 - 4 |Q_l| \es   +8 Q^2_l s_{eff}^4 \left[  1 + (Im\  \hat{k}_l)^2 \right] \, .
\fin
In the case of the four lightest quarks ($f = u - c$), Eq.~(\ref{e13}) is
replaced by
\begin{eqnarray} 
  1 - 4 |Q_f| \sms Re\ \hat{k}_f +8 Q^2_f \hat{s}^4 | \hat{k}_f |^2  =
 1 - 4 |Q_f| \es \left[1 + Re(\Delta \hat{k}_f - \Delta \hat{k}_l)
\right] \nonumber \\  \label{extra}
+  8 Q^2_f s_{eff}^4 \left[  1 + 
2\,Re(\Delta \hat{k}_f - \Delta \hat{k}_l)+ (Im\  \hat{k}_f)^2 \right] \, .
\end{eqnarray}
Next, we turn our attention to the overall coupling $(\ems / \sms \cms )$ and
the correction factor $\hat{\eta}_f = 1 + (\ems/\sms) \Delta \hat{\eta}_f$ in Eq.~(\ref{e11}).
It is convenient to split the latter into one- and two-loop contributions:
\ini \label{e14}
(\ems/\sms) \Delta \hat{\eta}_f = (\ems/\sms) \Delta \hat{\eta}_f^{(1)} +  
(\ems/\sms)^2 \Delta \hat{\eta}_f^{(2)} \, .
\fin
The amplitudes $(\ems/\sms) \Delta \hat{\eta}_f^{(1)}$ and  
$(\ems/\sms)^2 \Delta \hat{\eta}_f^{(2)}$ involve the field renormalization
of the $Z$ boson and appropriate vertex contributions and have been
studied in Ref.~\cite{c6} and Ref.~\cite{c13}, respectively\footnote{In Ref.~\cite{c6},
 $1 + (\ems/\sms) \Delta \hat{\eta}_f^{(1)}$ is denoted as $\overline{\rho_{ff}}$.}.

 In order to express $(\ems / \sms \cms )$ in terms of $G_\mu M_Z^2$, we employ
 \ini \label{e15}
\frac{\ems}{\sms } = 8 \frac{G_\mu}{\sqrt{2}}
M_W^2 \left(1 - \frac{\ems}{\sms}\hat{f}\right) \, ,  
 \fin
which follows from Eq.~(\ref{e8}) of Ref.~\cite{c11}. The amplitude $(\ems/\sms) \hat{f}$
involves the $W$-boson self-energy, as well as vertex and box contributions to
muon decay, and is discussed in that work. Recalling the relations \cite{c1}, \cite{c5}
\ini \label{e16}
\frac{M_W^2}{M_Z^2} = \cos^2 {\theta}_W \equiv c^2 \, ,
\fin
\ini \label{e17}
 \frac{c^2}{\cms} = \frac{1}{1 -(\ems/\sms)\Delta \hat{\rho}}\, ,  
\fin
the overall coupling may be expressed in the form:
\ini \label{e18}
\frac{\ems}{\sms \cms} = 8 \frac{G_\mu}{\sqrt{2}}
\frac{M_Z^2 \left(1 - (\ems/\sms)\hat{f}\right)}
{\left(1 - (\ems/\sms)\Delta \hat{\rho}\right)} \, .    
\fin
In analogy with Eq.~(\ref{e14}), it is convenient to split $(\ems/\sms) \Delta \hat{\rho}$
and $(\ems/\sms) \hat{f}$ into one and two-loop contributions. Noting that
$\Delta \hat{\eta}_f^{(1)}$ and $\hat{f}^{(1)}$ do not contain contributions
of ${\cal O}\left( (M_t^2/M_W^2)^n \right)$ $(n=1,2)$, and expanding in powers of
$G_\mu M_Z^2$, we find
\begin{eqnarray} \label{e19} \nonumber  
\frac{\ems}{\sms \cms} \, \hat{\eta}_f &=& \lambda \left\{ 1 +  \lambda c^2 
\left[ \Delta \hat{\rho}^{(1)} + \Delta \hat{\eta}_f^{(1)} - \hat{f}^{(1)} \right] 
\right.\\ \nonumber
 &+&  \lambda^2 c_{eff}^4  
\left[ \Delta \hat{\rho}^{(2)} + \Delta \hat{\eta}_f^{(2)} - \hat{f}^{(2)} \right.\\ 
 &+& 
\left. \left.(\Delta \hat{\rho})_{lead} 
\left( 2 \Delta \hat{\rho}^{(1)} - (\Delta \hat{\rho})_{lead}  
+ \Delta \hat{\eta}_f^{(1)} - 2 \hat{f}^{(1)}  \right) \right] \right\} \, ,
\end{eqnarray}
where 
\ini
\lambda \equiv \frac{8 G_\mu M_Z^2}{\sqrt{2}}
\fin
 and 
\ini
\left( \Delta \hat{\rho}\right)_{lead} = \frac{3}{64 \pi^2} \frac{M_t^2}{M_W^2}\fin
is the leading contribution to $\Delta \hat{\rho}^{(1)}$. Since 
$\left(\Delta \hat{\rho}\right)_{lead}$ does not involve $\sms$, and
$\Delta \hat{\eta}_f^{(1)}$ and $\hat{f}^{(1)}$ do not contain terms proportional
to $M_t^2/M_W^2$, we replace 
\mbox{$\sms \to \es$} everywhere in Eq.~(\ref{e19}).
In fact, the difference is a two-loop level effect not enhanced by powers of
$M_t^2/M_W^2$. The functions $\Delta \hat{\rho}^{(i)}$, $\Delta \hat{\eta}_f^{(i)}$,
$\hat{f}^{(i)}$ $(i = 1,2)$ still depend on both $c_{eff}^2$ 
and $c^2$. Furthermore,
the one-loop correction in Eq.~(\ref{e19}) is proportional to $c^2$. As explained in
Ref.~\cite{c8}, in order to ensure the complete cancellation of the scale dependence,
it is important to express the results in terms of a unique version of the
electroweak mixing angle. In order to achieve this objective, we employ the
strategy explained in Ref.~\cite{c8}: i) we replace everywhere $M_W^2$ by
$M_Z^2 c^2$; ii) in the two-loop contributions we substitute $c^2 \to c^2_{eff}$,
since the induced difference is of third order; iii) in the one loop
expressions we perform a Taylor expansion about $c^2 = c^2_{eff}$, using
the one-loop expression for  $c^2 - c^2_{eff}$ (cf.
 Eqs.~(9,10) of Ref.~\cite{c8}).
In practice, it is advantageous to evaluate the corresponding 
derivative numerically,
rather than analytically.

Furthermore, since
\begin{displaymath}
c^2 \left( \Delta \hat{\rho}\right)_{lead} = \frac{3}{64 \pi^2} \frac{M_t^2}{M_Z^2} 
\end{displaymath}
is independent of $c^2$ and the remaining one-loop contributions in Eq.~(\ref{e19}),
namely
\begin{displaymath}
\lambda c^2 
\left[ \Delta \hat{\rho}^{(1)}- (\Delta \hat{\rho})_{lead} + \Delta \hat{\eta}_f^{(1)} 
- \hat{f}^{(1)} \right] \, ,
\end{displaymath}
are not enhanced by $(M_t^2/M_W^2)^n $ $(n=1,2)$, it suffices
to replace $c^2 - c^2_{eff}$ by its leading contribution $c^2_{eff} x_t$
$(x_t = 3 G_\mu M_t^2 / \sqrt{2} \, 8  \pi^2)$. Substituting Eq.~(\ref{e13}) 
and the result derived
from Eq.~(\ref{e19}) into Eq.~(\ref{e11}), we obtain the final expression for 
$\Gamma_l$
in the EFF scheme.

\boldmath
\section{Simple formulae for  $\sef$, $M_W$, and $\Gamma_l$}
\unboldmath

In this Section we present simple analytic formulae for  $\es$, $M_W$, and
$\Gamma_l$
that reproduce accurately 
the results of the calculations reported in Refs.~\cite{c8,c11,c13},
and Section 2 of the present paper,
 over the range
$20\, \mbox{GeV} \le M_H \le 300\, \mbox{GeV}$. We have extended considerably the lower bound
of the range since current $M_W$ measurements, by themselves, 
favor low values of $M_H$.

The formulae are of the form
\begin{eqnarray} \label{e1}
\es & = & (\es)_0 + c_1 A_1 + c_5 A_1^2 + c_2 A_2 - c_3 A_3 + c_4 A_4, \\
\label{e2}
M_W & = & M_W^0 - d_1 A_1 - d_5 A_1^2 - d_2 A_2 + d_3 A_3 - d_4 A_4, \\
\label{GammaZ}
\Gamma_l & = & \Gamma_l^0 - g_1\,A_1 - g_5\,A_1^2  - g_2\,A_2 + g_3\,A_3 - 
g_4\,A_4 \, , 
\end{eqnarray}
where  
\begin{eqnarray}
A_1 \equiv \ln \left( M_H / 100 \, \mbox{GeV}\right) \, ,\quad &
A_2 \equiv \left[ \Delta \alpha_h^{(5)}/0.02761 \right]  - 1 \, , \nonumber \\
A_3 \equiv \left( M_t / 174.3 \, \mbox{GeV}\right)^2 -1 \, , \quad &
A_4 \equiv \left[ \alpha_s(M_Z) / 0.118 \right] -1 \, . 
\end{eqnarray}
We employ the input parameters \cite{leppages,us,dah1}
\begin{eqnarray}
M_Z = 91.1875 \, \mbox{GeV} \, , \quad & \alpha = 1/137.03599976\, , \nonumber \\
G_\mu = 1.16637 \times 10^{-5} \, \mbox{GeV}^{-2} \, , \quad &
\Delta \alpha_h^{(5)} = 0.02761 \pm 0.00036 \, , \nonumber \\ \label{inp}
M_t = 174.3 \pm 5.1 \, \mbox{GeV} \, , \quad &
\alpha_s \left( M_Z\right) = 0.118 \pm 0.002 \, ,
\end{eqnarray}
and include QCD corrections in the $\mu_t = \hat{m} \left( \mu_t\right)$
approach explained in Refs. \cite{c11,c14}. In contrast with the expressions
for $\es$ given in Ref.~\cite{c12}, we note that Eq.~(\ref{e1}) contains a 
term quadratic in $A_1$, a modification we have introduced in order to
 extend the accuracy of the formula to low $M_H$ values.
Eqs.~(\ref{e1}, \ref{e2}, \ref{GammaZ})
 incorporate complete one-loop corrections, as well as
two-loop contributions proportional to $(M_t^2/M_W^2)^n$ ($n = 1,2$), studied 
in Refs. \cite{c10,c11,c13,c15}, and Section 2 of the present paper. 

Tables \ref{t1}, \ref{t2}, and \ref{t5}
 present the constants $(\es)_0$, $M_W^0 $,
$\Gamma_l^0$,
$c_i$, $d_i $,
$g_i$ \mbox{($i=1 -5$)}  in the $\overline{MS}$ , OSI, OSII, and EFF schemes. 
In particular, OSII is strictly independent of the 
electroweak scale, a property that it shares with the EFF scheme. Over the 
range $20\, \mbox{GeV} \le M_H \le 300 \, \mbox{GeV}$, and varying the input parameters
within $1 \sigma$, these formulae reproduce the calculations from the 
detailed codes with maximum absolute deviations $\approx 10^{-5}$ in $\es$,
$\approx 0.8 \, \mbox{MeV}$ in $M_W$, and  $\approx 7 \, \mbox{KeV}$ in $\Gamma_l$.

Although we have employed $\dah = 0.02761 \pm 0.00036$ \cite{dah1} in the 
determination of $(\es)_0$, $M_W^0$, $\Gamma_l^0$, $c_i$, 
$d_i$, $g_i$  $(i = 1 - 5 )$, we recall
that recent calculations of this important quantity range from
$0.027426 \pm 0.000190$ \cite{dah2} to $0.027896 \pm 0.000395$
\cite{dah3}.
For this reason, we have examined the validity of Eqs.~(\ref{e1}, \ref{e2},
\ref{GammaZ}),
with the coefficients given in Tables \ref{t1}, \ref{t2}, and \ref{t5},
 over the 
large range
$0.0272 \le \dah \le 0.0283$ that encompasses the recent calculations.
We find that the simple formulae still reproduce the results from the
detailed codes with maximum absolute deviations that are very close to
those reported above.

It is worthwhile to note that the current
experimental uncertainty in $G_\mu M_Z^2$ ($\pm 0.0047 \%$) induces tree-level
errors of $1.6 \times 10^{-5}$ in $\es$, $0.78 \, \mbox{MeV}$ in $M_W$,
and $ 4 \, \mbox{KeV}$ in $\Gamma_l$,
which are of the same magnitude as the deviations discussed above.
Thus, the accuracy within which 
Eqs.~(\ref{e1},
\ref{e2}, \ref{GammaZ}) reproduce the calculations of the detailed codes
employed in  Refs. \cite{c8,c11,c13}
and of the present paper, is sufficient
 at 
present.

As a first application, we employ Eq.~(\ref{e1}) in the EFF scheme to obtain
the value of $M_H$ derived from the current world average $(\es)_{exp} = $
$ 0.23149 \pm
0.00017$. Noting that $x \equiv c_1 A_1 + c_5 A_1^2$ is normally distributed,
we obtain $x = (1.07 \pm 2.67) \times 10^{-4}$, $x < 5.46 \times 10 ^{-4}$
at the $95 \, \%$ confidence level, which leads to 
\ini \label{e5}
M_H = 124_{-52}^{+82}\, \mbox{GeV};\quad M_H^{95} = 280 \, \mbox{GeV} ,
\fin
where $M_H^{95}$ stands for the  $95 \, \% $ CL upper bound.

We may also determine $M_H$ using the experimental value of $M_W$ in 
conjunction with Eq.~(\ref{e2}). The current average of the $p \overline{p}$ -
collider and LEP2 measurements is $(M_W)_{exp} = 80.451 \pm 0.033 \, \mbox{GeV}$ 
\cite{leppages}.
Inserting this value in Eq.~(\ref{e2}) and calling
$ y \equiv d_1 A_1 +d_5 A_1^2$, we obtain in the EFF scheme 
$ y = -0.0648 \pm 0.0470$, $y < 0.0117 $ at $95 \, \%$ CL , 
which corresponds to 
\ini \label{e7}
M_H = 23_{-23}^{+49}\, \mbox{GeV}\,;\quad  M_H^{95} = 122 \, \mbox{GeV}\, .
\fin
The results given in Eqs.~(\ref{e5},\ref{e7}) have been 
confirmed on the basis
of a \mbox{$\chi^2$-analysis} based on the simple formulae. The central values 
of $M_H$ have also been verified using the detailed codes.

In summary, for current
experimental inputs, $M_W$ leads to significantly lower values of $M_H$ than
those derived from $\es$. This divergence can be also illustrated 
by using the $A_1$ value derived from $\es = 0.23149 \pm 0.00017$ 
and Eq.~(\ref{e1}), to predict $M_W$ via Eq.~(\ref{e2}).
Taking into account the correlation of errors in Eqs.~(\ref{e1}, \ref{e2}),
one finds 
\ini \label{e8}
\left( M_W \right)_{indir.} = \left( 80.374 \pm 0.025 \right)\, \mbox{GeV}\, ,
\fin
which differs from $(M_W)_{exp.}$ by $1.86\,\sigma$. 
It is worth noting that the indirect determination
derived by using $\es$ (Eq.~(\ref{e8})) is quite close to the 
corresponding value $(M_W)_{indir.} = 80.379 \pm 0.023\, \mbox{GeV}$ obtained
in the global analysis \cite{leppages}.

It is also very interesting to obtain $M_H$ using simultaneously
the experimental values of $\es$, $M_W$, and $\Gamma_l$. 
Employing $(\es)_{exp}$ and $(M_W)_{exp}$, and the current world average 
$(\Gamma_l)_{exp} = 83.984 
\pm 0.086 \, \mbox{MeV}$, a $\chi^2$ analysis based on the simple
formulae in the EFF scheme leads to
\ini \label{eq21}
M_H = 97_{-41}^{+66}\, \mbox{GeV}\,;\quad  M_H^{95} = 223 \, \mbox{GeV}\, .
\fin
This result is of course dominated by the $\es$ and $M_W$ data, and may be 
compared with  $M_H^{95} = 196 \, \mbox{GeV}$ in the global 
fit \cite{leppages}. The difference is partly due to the fact that 
$(M_t)_{fit}$ turns out to be somewhat smaller in the global analysis than 
in our case.

We remind the reader that the current determination of $(\es)_{exp}$
shows a $\chi^2/\mbox{d.o.f.}=10.6/5$, which corresponds to a confidence level
of only $6\%$. Furthermore, the experimental results exhibit an intriguing
dichotomy. Those based on the leptonic observables ($A_l(SLD)$, $A_l(P_\tau)$,
$A_{fb}^{(0,l)}$) lead to  $(\es)_{l}=0.23113 \pm 0.00021$
while those derived from the hadronic sector ($A_{fb}^{(0,b)}$, $A_{fb}^{(0,c)}$,
$<Q_{fb}>$) have an average $(\es)_{h}=0.23220 \pm 0.00029$. The results within each
group agree well with each other but the averages of the two sectors
differ by about $3.0 \sigma$!

Using Eq.~(\ref{e1}) in the EFF scheme, we find that  $(\es)_{l}$ leads to
\ini
M_H = 59_{-29}^{+50}\, \mbox{GeV}\,;\quad  M_H^{95} = 158 \, \mbox{GeV}\, ,
\fin
a result significantly closer to that 
derived from $(M_W)_{exp}$ (Eq.~(\ref{e7}))!
The intriguing question remains of whether the difference between
$(\es)_{l}$ and $(\es)_{h}$ reflects a statistical fluctuation or is due
to new physics involving perhaps the third generation of quarks. In
the case of a statistical fluctuation, a possible method of dealing
with the discrepancy is to enlarge the error, as discussed in 
Ref.~\cite{c12,gurtu}. For instance, if the $\es$ uncertainty is
increased by a factor $[\chi^2/\mbox{d.o.f.}]^{1/2}$ according
to the PDG prescription \cite{pdg}, we would have $\es = 0.23149 \pm 0.00025$
which, using Eq.~(\ref{e1}) in the EFF scheme leads to
\ini
M_H = 124_{-60}^{+104}\, \mbox{GeV}\,;\quad  M_H^{95} = 331 \, \mbox{GeV}\, ,
\fin
and 
\ini
M_H = 76_{-37}^{+64}\, \mbox{GeV}\,;\quad  M_H^{95} = 201 \, \mbox{GeV}\, ,
\fin
in the combined $\es$-$M_W$-$\Gamma_l$ analysis.
We note that the enlarged error leads to an increase of $M_H^{95}$
from $280\, \mbox{GeV}$ (Eq.~(\ref{e5})) to $331\, \mbox{GeV}$ if we only
employed $\es$ as an input, and to a decrease 
from  $223\, \mbox{GeV}$ (Eq.~(\ref{eq21}))
to $201\, \mbox{GeV}$ in the combined $\es$-$M_W$-$\Gamma_l$ fit.
The latter result may be understood as follows: the increasing
error in $\es$ changes the shape of the $\chi^2 (M_H)$ curve and shifts
the position of its minimum to a smaller $M_H$ value.
In our case the combination of these two effects leads to a decrease in 
$M_H^{95}$.
Although this scaling method is not 
generally employed in current analysis of the
electroweak data, it provides a more conservative estimate of $\es$.
It is very interesting to note that it has a small effect in the determination
of $M_H^{95}$, particularly in analyses that combine the $\es$ and $M_W$
data, and that it leads in such a case to a decrease in  $M_H^{95}$!
If the discrepancy is due to new physics involving the third generation of
quarks \cite{marci,chano,tenyears} the analysis shows that a substantial,
tree-level change in the right-handed $Zb\overline{b}$ coupling is required.
Very recently, it has also been argued that, if the $(\es)_{h}$ and $(\es)_{l}$
discrepancy were to settle on the leptonic side, a ``new physics'' scenario involving
light sneutrinos and gauginos would substantially improve 
the agreement with the 
electroweak data and the direct lower bound on $M_H$ \cite{altapaolo}.

\boldmath
\section{Comparative analysis of the OS, $\overline{MS}$, and EFF 
renormalization schemes} 
\unboldmath

As mentioned in the introduction, because of the truncation of the 
perturbative series, the $\msbar$ calculations have a residual scale 
dependence, while the EFF scheme is strictly scale independent.
This is illustrated in Fig.~1, which compares, in the case
$M_H = 100 \, \mbox{GeV}$, the 
scale dependence of $\Gamma_l$ in the $\msbar$ scheme with the
scale-independent value in the EFF scheme.
We note that the latter coincides, to very good approximation, with the maximum
of the $\msbar$ curve, which occurs at $\mu \approx 75 \,\mbox{GeV}$, a scale 
somewhat smaller than $M_Z$.
Nonetheless, the difference between the EFF value and the $\msbar$
calculation at $\mu = M_Z$ is very small.

Tables \ref{t8}, \ref{t9}, \ref{t22}, present the value of $\es$, $M_W$, and
$\Gamma_l$, as functions of $M_H$, in the four schemes discussed in the paper.
They have been obtained using the detailed numerical codes employed in 
Refs.~\cite{c8,c11,c13} and the present paper, the central values of the input
parameters in Eq.~(\ref{inp}),
and $\mu =M_Z$  in the case of the $\msbar$ and OSI schemes.
We have kept six significant figures to show more precisely the differences 
between the various schemes.

It is interesting to note that, in contrast with $\es$ and $M_W$, $\Gamma_l$
is not a monotonic function of $M_H$. Indeed, it shows a maximum at 
$M_H \approx 52\, \mbox{GeV}$.

Examination of Table~\ref{t8} shows that the maximum difference for
$\es$ among the four schemes amounts to $3.7 \times 10^{-5}$ and occurs between
OSII and EFF at $M_H = 200, 250, 300 \, \mbox{GeV}$. For $M_W$, the maximum difference
is $2.4 \, \mbox{MeV}$ and occurs between $\msbar$ and OSI at $M_H = 20 \, \mbox{GeV}$.
Finally, for $\Gamma_l$, the maximum difference is $3.3 \, \mbox{KeV}$ and occurs 
between OSI and OSII at $M_H = 300 \, \mbox{GeV}$. 
These differences are quite similar to those encountered in Ref.~\cite{c8,c11}
for the $\es$ and $M_W$ calculations over the range $65\, \mbox{GeV} \le M_H
\le 600 \, \mbox{GeV}$. They illustrate the scheme dependence of current 
calculations and provide a first estimate of the  theoretical
error due to the truncation of the perturbative series. Taking into account 
further uncertainties associated with the QCD corrections \cite{c12,phili}, 
the theoretical errors may be enlarged to
$\approx 6 \times 10^{-5}$ in $\es$ and $\approx 7 \, \mbox{MeV}$ in $M_W$.

In the case of $M_W$, an alternative comparison has been made between OSII, 
with an $M_t$-implementation of QCD corrections \cite{c11}, and 
important recent
 calculations that include all ${\cal O}(\alpha^2)$ contributions to 
$\Delta r$ that contain a fermionic loop \cite{hollik}.
The latter incorporate not only the terms of order ${\cal O}(\alpha^2 (M_t^2 /
M_W^2)^n)$ ($n=1, 2$), but also a subset of ${\cal O}(\alpha^2)$ diagrams 
not enhanced by powers of $M_t^2/M_W^2$. Over the range $65\, \mbox{GeV} \le M_H
\le 600 \, \mbox{GeV}$, Ref.~\cite{hollik} reported a monotonically decreasing 
difference between OSII and the more complete calculations, amounting
to $\approx 5 \, \mbox{MeV}$ at $M_H = 100 \, \mbox{GeV}$
when the two-loop leptonic contributions to $\Delta \alpha$ are included
in both calculations.
A glance at Table~\ref{t9} tells us then that the difference between the other
schemes and Ref.~\cite{hollik} is somewhat larger, reaching $\approx 7 
\, \mbox{MeV}$ for $\msbar$ at $M_H = 100 \, \mbox{GeV}$.

Since the more complete calculations have not been implemented so far in 
the case of $\es$ and $\Gamma_l$, similar estimates of the theoretical 
error for these important amplitudes are not yet available. It has been argued 
\cite{hollik} 
that the $M_W$ difference between OSII and the more complete calculations may be used to 
provide an estimate of the theoretical error in the evaluation of $\es$, via 
the OS relation
\ini \label{e21}
\es = (1 + \Delta k)\, s^2 \, ,
\fin 
where $s^2 = \sin^2 \theta_W = 1 - M_W^2 /M_Z^2$ and $1 + \Delta k$ is the 
relevant electroweak form factor. In this approach, the uncertainty in $M_W$,
and therefore in $s^2$, is translated into an estimate of the error in the 
calculation of $\es$. 
In our view, such arguments may well be misleading since important corrections
involving
\ini
X = \frac{c^2}{s^2}\,  Re \left[ \frac{A_{WW}(M_W^2)}{M_W^2} -
\frac{A_{ZZ}(M_Z^2)}{M_Z^2}\right]
\fin
where $A_{WW}$ and $A_{ZZ}$ are the $W$ and $Z$ self-energies, contribute with
opposite signs to $\Delta k$ and $\Delta r$ \cite{c1}.
As a consequence,  an additional contribution $\delta X$ to $X$ 
leads to $\delta s^2 = - [s^2 c^2 /(c^2 -s^2)]\delta X$, but only
 $\delta \es = - [s^4 /(c^2 -s^2)]\delta X$! Thus, $\Delta k$ induces
a suppression factor $s^2/c^2$ in this important class of contributions
to $\es$.
The fact that the theoretical uncertainty in 
$\es$ cannot be simply related to the error in the evaluation of $M_W$ may 
also be understood by recalling salient features of the $\msbar$ calculation 
of $\es$. In the $\msbar$ context, $(1 + \Delta k)\,s^2$ in Eq.~(\ref{e21}) 
is replaced by $(1 + \Delta \hat{k})\,\hat{s}^2$, an expression
that does not explicitly involve $M_W$ at the three level! This feature is 
even more striking in the EFF calculation of $\es$, where $M_W$ completely 
decouples \cite{c8}.
It is also worth noting that, in the $\msbar$ calculation of $\es$, the 
dependence on $X$ does not involve the enhancement factor $c^2/s^2$. One 
readily finds that an additional correction $\delta X$ contributes
$\approx -[\hat{s}^4 / (\hat{c}^2 - \hat{s}^2)] \, \delta X$
to $(1+ \Delta \hat{k}) \hat{s}^2$, a result analogous to that found in the OS
framework!

\section{Conclusions}

In Section 2, we have discussed the calculation of the leptonic partial
width $\Gamma_l$ of the $Z$ boson in the effective scheme of renormalization.
The extension to the case of the four lightest quarks involves simple 
modifications, outlined in the text. Our starting
point has been the corresponding expressions in the $\msbar$ 
approach \cite{c13}. Both formulations include the corrections of 
${\cal O}\left( g^4 (M_t^2/M_W^2)^n \right)$ $(n=1,2)$.
A salient feature of the calculations in the effective scheme is that,
in contrast with their $\msbar$ counterparts,  they 
are strictly scale independent in finite orders of perturbation
theory. 
This is illustrated in Fig.~1 which depicts, for 
$M_H = 100 \, \mbox{GeV}$, the scale dependence 
of current calculations
of $\Gamma_l$ in the $\msbar$
approach and its comparison with the effective scheme result.

In Section 3, we have presented very simple formulae for 
$\sef$, $M_W$, and $\Gamma_l$, as functions of $M_H$ and important input 
parameters such as $\dah$, $M_t$, and $\asmz$. A novel feature
of these formulae is that their range of validity has been extended to low
$M_H$ values probed by current experiments.
Furthermore, they retain their accuracy over the range of $\dah$
values spanned by recent calculations of this important amplitude.

We have illustrated the usefulness of these formulae by extracting
$M_H$ and its $95\%$ CL upper bound $M_H^{95}$ employing separately,
as inputs, the current experimental averages $(\es)_{exp}$ and 
 $(M_W)_{exp}$, as well as the combination of  $(\es)_{exp}$, $(M_W)_{exp}$,
and  $(\Gamma_l)_{exp}$. We have also
employed the formulae to predict $M_W$ using $(\es)_{exp}$ as an input.
The result of this prediction turns out to be quite close to that currently derived using the 
global analysis of all the available electroweak data.

We have also discussed the well-known discrepancy between  
$(\es)_{l}$ and $(\es)_{h}$, and some of its possible implications for new 
physics.
On the other hand, assuming that the difference is due to statistical 
fluctuations, we have also considered the effect of enlarging 
the error according to the well-known PDG prescription.
Surprisingly, we have found that such a procedure decreases $M_H^{95}$ by
a small amount in analyses that combine $\es$ and $M_W$ as inputs.

In Section 4, we have compared the calculations of $\sef$, $M_W$, and  
$\Gamma_l$ in the EFF, $\msbar$, OSI, and OSII schemes,
over the range $20\, \mbox{GeV} \le M_H \le 300\, \mbox{GeV}$.
We have found maximal differences of 
$3.7 \times 10^{-5}$ in $\sef$,  $2.4 \, \mbox{MeV}$ in $M_W$
and $3.3 \, \mbox{KeV}$ in  $\Gamma_l$.
Taking into account uncertainties in the QCD corrections \cite{c12,phili},
the theoretical errors may be enlarged to 
$\approx 6 \times 10^{-5}$ in $\es$ and $\approx 7 \, \mbox{MeV}$ in $M_W$.
In the case of $M_W$, an alternative comparison can
be made with important recent calculations that include all
${\cal O}(\alpha^2)$ contributions to $\Delta r$ that contain a
fermionic loop \cite{hollik}.
This reveals differences  $\approx 7 \, \mbox{MeV}$ at  
$M_H = 100\, \mbox{GeV}$ with the four schemes discussed in this paper.

Finally, we have discussed the theoretical shortcomings of estimates 
of the theoretical error in  $\sef$ derived solely from the
uncertainty of $M_W$ in the OS scheme.

\section*{Acknowledgments}
The authors are greatly indebted to G.~Degrassi and P.~Gambino for very 
valuable communications. This research was supported in part by the NSF
grant PHY-0070787. M.P.~acknowledges partial support from the European 
Program HPRN-CT-2000-00149.

\begin{figure}[h!] \label{gammafig}
\centering
\psfrag{m}{ {\small $\mu\ \left[ \mbox{GeV} \right]$}}
\psfrag{s}{{\footnotesize $\Gamma_l\, [\mbox{MeV}]$ }}
\resizebox{12cm}{7cm}{\includegraphics{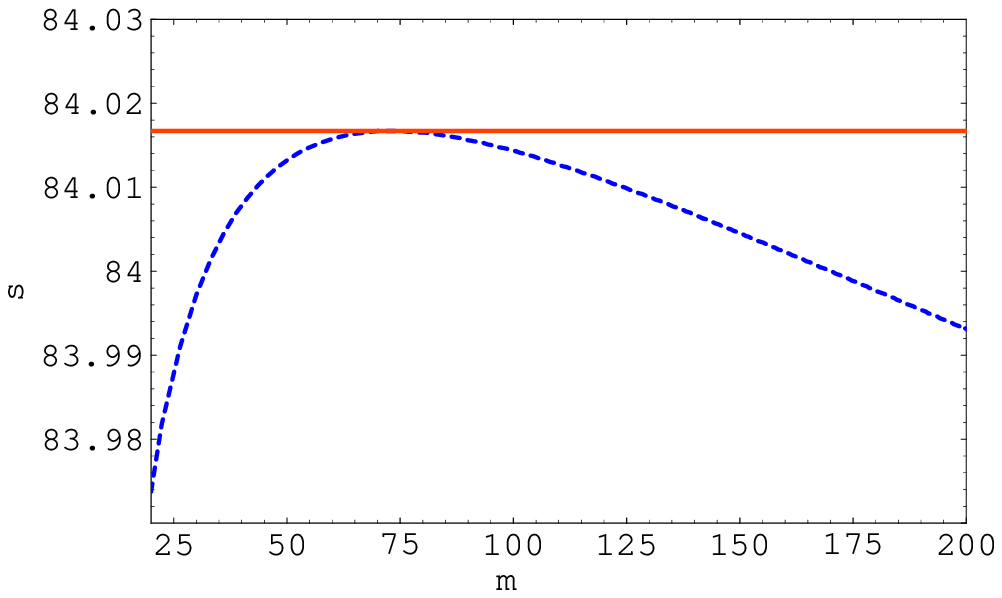}}
\caption{Scale dependence of $\Gamma_l$ in the $\overline{MS}$ (dashed line)
and EFF (solid line) schemes for $M_H =  100\, \mbox{GeV}$.}
\label{fig1}
\end{figure}


\begin{table}[hp] 
\caption{Values of the constants appearing in Eq.~(\ref{e1}).} 
\label{t1}
\begin{center}
\begin{tabular}{|c||c|c|c|c|c|c|}
\hline
 &  &  &  &  &  & \\
Scheme & $\left(s_{eff}^2 \right)_0$ & $10^4\, c_1$ & $10^3\, c_2$ &
 $10^3\, c_3$ &
$10^4\,  c_4$ &  $10^5\, c_5$ \\
&  &  &  &  &  & \\
\hline \hline
&  &  &  &  &  & \\
$\overline{MS}$ & $ 0.231392 $ & $ 4.908 $ & $ 9.69 $ & $ 2.77 $ &
$ 4.5 $ & $3.43$ \\ 
&  &  &  &  & & \\ \hline
&  &  &  &  & & \\
$OSI$ & $ 0.231405 $ & $ 4.877 $ & $ 9.69 $ & $ 2.76 $ &
$ 4.5 $ & $3.34$ \\ 
&  &  &  &  & & \\
\hline
&  &  &  &  & & \\
$OSII$ & $ 0.231417 $ & $ 4.999  $ & $ 9.69 $ & $ 2.69 $ &
$ 4.4 $ & $3.37$ \\ 
&  &  &  &  &  & \\
\hline
&  &  &  &  &  & \\
$EFF$ & $ 0.231383 $ & $ 4.948 $ & $ 9.69 $ & $ 2.78 $ &
$ 4.5 $ &$3.50 $  \\
&  &  &  &  &  &\\  
\hline
\end{tabular}
\end{center}
\end{table}

\begin{table}[hp] 
\caption{Values of the constants appearing in Eq.~(\ref{e2}), in $\mbox{GeV}$.} 
\label{t2}
\begin{center}
\begin{tabular}{|c||c|c|c|c|c|c|}
\hline
 &  &  &  &  &  & \\
Scheme & $M_W^0$ & $10^2 \, d_1$ & $10\, d_2$ & $10\, d_3$ &
$10^2\, d_4$ & $10^3\, d_5$ \\
&  &  &  &  &  &\\
\hline \hline
&  &  &  &  &  &\\
$\overline{MS}$ & $ 80.3868 $ & $ 5.719 $ & $ 5.07 $ & $ 5.42 $ &
$ 8.5 $ & $ 8.98  $\\ 
&  &  &  &  &  & \\ \hline
&  &  &  &  &  & \\
$OSI$ & $ 80.3849 $ & $ 5.667 $ & $ 5.08 $ & $ 5.40 $ &
$ 8.5 $ & $ 8.85  $\\ 
&  &  &  &  &  & \\
\hline
&  &  &  &  &  & \\
$OSII$ & $ 80.3847 $ & $ 5.738 $ & $ 5.08 $ & $ 5.37 $ &
$ 8.5 $  & $ 8.92 $\\ 
&  &  &  &  &  & \\
\hline
&  &  &  &  &  & \\
$EFF$ & $ 80.3862 $ & $ 5.730 $ & $ 5.08 $ & $ 5.42 $ &
$ 8.5 $ & $ 8.98 $\\
&  &  &  &  &  & \\  
\hline
\end{tabular}
\end{center}
\end{table}

\begin{table}[hp] 
\caption{Values of the constants appearing in Eq.~(\ref{GammaZ}), in $\mbox{MeV}$.} 
\label{t5}
\begin{center}
\begin{tabular}{|c||c|c|c|c|c|c|}
\hline
 &  &  &  &  &  & \\
Scheme & $\Gamma_l^0$ & $10^2 \, g_1$ & $10\, g_2$ & $10\, g_3$ &
$10\, g_4$ & $10^2\, g_5$ \\
&  &  &  &  &  &\\
\hline \hline
&  &  &  &  &  &\\
$\overline{MS}$ & $ 84.0164 $ & $ 4.439 $ & $ 4.77 $ & $ 8.03 $ &
$ 1.13 $ & $ 3.670  $\\ 
&  &  &  &  &  & \\ \hline
&  &  &  &  &  & \\
$OSI$ & $ 84.0174 $ & $ 4.472 $ & $ 4.76 $ & $ 8.04 $ &
$ 1.13 $ & $ 3.673  $\\ 
&  &  &  &  &  & \\
\hline
&  &  &  &  &  & \\
$OSII$ & $ 84.0152 $ & $ 4.568 $ & $ 4.76 $ & $ 8.01 $ &
$ 1.12 $  & $ 3.683 $\\ 
&  &  &  &  &  & \\
\hline
&  &  &  &  &  & \\
$EFF$ & $ 84.0176 $ & $ 4.549 $ & $ 4.77 $ & $ 8.05 $ &
$ 1.13 $ & $ 3.696 $\\
&  &  &  &  &  & \\  
\hline
\end{tabular}
\end{center}
\end{table}

\begin{table}[p] 
\caption{Values for $\es$ in the four renormalization schemes considered in 
the paper.
They have been calculated using the central values of the 
input parameters in Eq.~(\ref{inp}).}
\label{t8}
\begin{center}
\begin{tabular}{|c||c|c|c|c|}
\hline
&  &  &  & \\
$M_H\,(\mbox{GeV})$ & $EFF$ & $\overline{MS}$ & $OSII$  & $OSI$\\
&  &  &  & \\
\hline \hline
&  &  &  & \\
$20$ & $ 0.230686 $ & $ 0.230699  $ & $ 0.230708 $ & $0.230715$\\ 
&  &  &  & \\ \hline
&  &  &  & \\
$50$ & $ 0.231051 $ & $ 0.231062 $ & $ 0.231080 $ & $0.231077$\\
&  &  &  & \\
\hline
&  &  &  & \\
$100$ & $ 0.231385 $ & $ 0.231393 $ & $ 0.231418 $ & $0.231406$\\ 
&  &  &  & \\
\hline
&  &  &  & \\
$150$ & $ 0.231593 $ & $ 0.231599 $ & $ 0.231628 $ & $0.231611$\\
&  &  &  & \\  
\hline
&  &  &  & \\
$200$ & $ 0.231744 $ & $ 0.231749 $ & $ 0.231781 $ & $0.231760$\\
&  &  &  & \\  
\hline
&  &  &  & \\
$250$ & $ 0.231864 $ & $ 0.231868 $ & $ 0.231901 $ & $0.231878$\\
&  &  &  & \\  
\hline
&  &  &  & \\
$300$ & $ 0.231964 $ & $ 0.231967 $ & $ 0.232001 $ & $0.231976$\\
&  &  &  & \\  
\hline
\end{tabular}
\end{center}
\end{table}

\begin{table}[p] 
\caption{Values for $M_W$ (in $\mbox{GeV}$) 
in the four renormalization schemes considered in 
the paper (calculated as in Table 4).}
\label{t9}
\begin{center}
\begin{tabular}{|c||c|c|c|c|}
\hline
&  &  &  & \\
$M_H\,(\mbox{GeV})$ & $EFF$ & $\overline{MS}$ & $OSII$  & $OSI$\\
&  &  &  & \\
\hline \hline
&  &  &  & \\
$20$ & $80.4550 $ & $ 80.4555  $ & $ 80.4539 $ & $80.4531$\\ 
&  &  &  & \\ \hline
&  &  &  & \\
$50$ & $ 80.4219 $ & $ 80.4224 $ & $ 80.4206 $ & $80.4202$\\
&  &  &  & \\
\hline
&  &  &  & \\
$100$ & $ 80.3863 $ & $ 80.3869 $ & $ 80.3849 $ & $80.3850$\\ 
&  &  &  & \\
\hline
&  &  &  & \\
$150$ & $ 80.3614 $ & $ 80.3621 $ & $ 80.3600 $ & $80.3604$\\
&  &  &  & \\  
\hline
&  &  &  & \\
$200$ & $ 80.3421 $ & $ 80.3428 $ & $ 80.3406 $ & $80.3413$\\
&  &  &  & \\  
\hline
&  &  &  & \\
$250$ & $ 80.3263 $ & $ 80.3270 $ & $ 80.3248 $ & $80.3256$\\
&  &  &  & \\  
\hline
&  &  &  & \\
$300$ & $ 80.3129 $ & $ 80.3135 $ & $ 80.3114 $ & $80.3124$\\
&  &  &  & \\  
\hline
\end{tabular}
\end{center}
\end{table}

\begin{table}[p] 
\caption{Values for $\Gamma_l$ (in $\mbox{MeV}$) 
in the four renormalization schemes considered in 
the paper (calculated as in Table 4).
}
\label{t22}
\begin{center}
\begin{tabular}{|c||c|c|c|c|}
\hline
&  &  &  & \\
$M_H\,(\mbox{GeV})$ & $EFF$ & $\overline{MS}$ & $OSII$  & $OSI$\\
&  &  &  & \\
\hline \hline
&  &  &  & \\
$20$ & $83.9896 $ & $ 83.9873  $ & $ 83.9878 $ & $83.9887$\\ 
&  &  &  & \\ \hline
&  &  &  & \\
$50$ & $ 84.0353 $ & $ 84.0335 $ & $ 84.0330 $ & $84.0346$\\
&  &  &  & \\
\hline
&  &  &  & \\
$100$ & $ 84.0167 $ & $ 84.0155 $ & $ 84.0143 $ & $84.0165$\\ 
&  &  &  & \\
\hline
&  &  &  & \\
$150$ & $ 83.9907 $ & $ 83.9900 $ & $ 83.9882 $ & $83.9908$\\
&  &  &  & \\  
\hline
&  &  &  & \\
$200$ & $ 83.9667 $ & $ 83.9664 $ & $ 83.9643 $ & $83.9672$\\
&  &  &  & \\  
\hline
&  &  &  & \\
$250$ & $ 83.9457 $ & $ 83.9457 $ & $ 83.9432 $ & $83.9463$\\
&  &  &  & \\  
\hline
&  &  &  & \\
$300$ & $ 83.9271 $ & $ 83.9274 $ & $ 83.9247 $ & $83.9280$\\
&  &  &  & \\  
\hline
\end{tabular}
\end{center}
\end{table}

\end{document}